\documentclass[aps,prb,twocolumn,superscriptaddress]{revtex4-1}

\usepackage{amsmath}
\usepackage{amssymb}
\usepackage{graphicx}
\usepackage{color}

\begin{document}

\preprint{LA-UR-09-06102}

\title{Electron-phonon coupling in semimetals in a high magnetic field}

\author{P. B. Littlewood}
\affiliation{Theory of Condensed Matter Group, Cavendish Laboratory, University of Cambridge,
             JJ Thomson Ave, Cambridge CB3 0HE, UK}
\author{B. Mihaila}
\affiliation{Los Alamos National Laboratory, Los Alamos NM 87545, USA}
\author{R. C. Albers}
\affiliation{Los Alamos National Laboratory, Los Alamos NM 87545, USA}

\date{\today}

\begin{abstract}
We consider the effect of electron-phonon coupling in semimetals in high magnetic fields, with regard to elastic modes that can lead to a redistribution of carriers between pockets. We show that in a clean three dimensional system, at each Landau level crossing, this leads to a discontinuity in the magnetostriction, and a divergent contribution to the elastic modulus. We estimate the magnitude of this effect in the group V semimetal Bismuth.
\end{abstract}

\pacs{72.15.Jf;63.20.kd; 71.55.Ak; 73.43.-f}

\maketitle

In a quantizing magnetic field, the electronic density of states is split into Landau levels, and in a clean three-dimensional system, there is a square-root singularity in the density of states as a function of energy. This is the origin of a host of magneto-oscillation effects, that {\em inter alia} provide a detailed tomography of fermi surfaces in metals. There are effects which are peculiar to multi-band rather than single-band systems. In a system with a single species of carrier, the chemical potential oscillates with field in order to keep the carrier number fixed. But with multiple pockets of carriers, there is an opportunity to redistribute carriers between pockets to minimize the energy with coupling generated either by electron-electron or electron-phonon coupling. In this paper, we concentrate on the latter, and especially coupling to uniform strain in the crystal.

The study of large magnetostriction in semimetals goes back to Kapitza~\cite{kapitza} and Shoenberg~\cite{shoenberg}, with work from the 1960's onwards concentrating on quantizing fields~\cite{chandrasekhar,fawcett,michenaud} up to 20~T. Bismuth (Bi) has recently come to the fore in a series of experiments in higher field~\cite{behnia07,ong,behnia09} -- the extreme quantum limit where only one or a few Landau levels are occupied -- which have produced puzzling results that we will discuss in more detail below. Part of the purpose of this paper is to point out the extent to which electron-phonon coupling should be included in any analysis, and to detail which of the phenomena observed experimentally {\em cannot} be produced by magnetostrictive effects, in order that the arena to search for new physics is clear.

We begin by a simple formulation of the problem and recapitulate general results. We then estimate the magnitude of the effects in \textit{Bi}, and discuss this in the context of both older and more recent results. Since the deformation potential coupling at the band edge is not well established experimentally, we use ab-initio calculations of the band edges under strain to guide our estimates.

\begin{figure}[b]
   \centering
   \includegraphics[width=0.95\columnwidth]{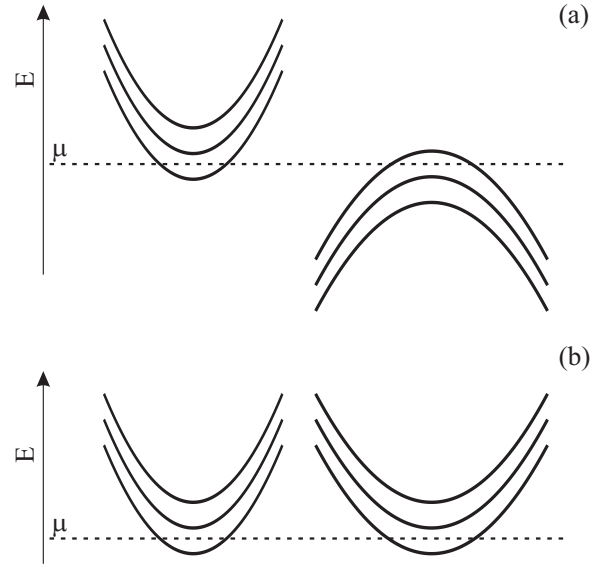}
\caption{\label{fig1} Sketch of the Landau levels of an electron and hole, where the chemical potential is indicated by a dashed line. The lower figure represents hole energies as positive, the convention used here.}
\end{figure}

For an illustrative discussion of the effects, we will consider the case of two pockets because no particular extra physics is introduced by further generalization. The Landau levels are then indexed by quantum number $n$ and momentum $k$ in the direction of magnetic field; again for simplicity we will assume the field to be oriented along a principal axis of the effective mass tensor. The carrier energies in the two pockets $i=1,2$ are
\begin{equation}
   E_i(n,k) = \epsilon_i + \bigl ( n + 1/2 \bigr ) \, \hbar \omega_{ic}
              + \frac{\hbar^2 k^2}{2 m_{iz}} \>,
\end{equation}
where $m_i$ is the effective mass, $\omega_{ic} = eB/m_{ic}c$ the cyclotron frequency (see Fig.~\ref{fig1}). Holes are treated as having positive energies. A chemical potential $\mu$ is introduced to fix the carrier densities $n_i$, and the band edges $\epsilon_i \rightarrow \epsilon_i+ \gamma_{i,\alpha} s_\alpha$ shift with strain $s_\alpha$. While it is true that other parameters -- for example the effective mass -- are strain-dependent, it is only the band-edge shift that contributes in a singular fashion, as seen below.

We specify two cases of interest:
\begin{enumerate}
   \item Electron and hole pockets: $n_1 = n_2$ constrains equal numbers in the two pockets. Here, $\epsilon_i \rightarrow \epsilon_i + \gamma s$ adjusts the (equal) numbers of electrons and holes so that the strain is emptying electrons from one pocket to the other.
   \item Two degenerate electron pockets: $n_1 + n_2 = n$, with the total carrier number fixed. Here, $\epsilon_{1,2} \rightarrow \epsilon_{1,2} \pm \gamma s$, so that the strain breaks the symmetry between the pockets.
\end{enumerate}
Of course, the type of strains in the two cases will be different.


We now compute, conventionally, the total electronic energy and the total occupancy. In each Landau level, the density of states (per unit energy, per unit area) is
\begin{equation}
   g(E) = \frac{1}{2 \pi^2 \ell^2} \left ( \frac{2m_{iz}}{\hbar^2}\right ) ^{1/2} E^{-1/2}\; ,
\end{equation}
and, therefore, the carrier density and the total electronic energy in each pocket become
\begin{eqnarray}
   n_i = \frac{1}{\pi^2 \ell^2} \left ( \frac{2 m_{iz}}{\hbar^2} \right ) ^{1/2} \sum_n \, (\mu - \epsilon_{ni})^{1/2} \, \Theta (\mu - \epsilon_{ni})\; ,
   \\
   E_i = \frac{1}{3 \pi^2 \ell^2} \left ( \frac{2 m_{iz}}{\hbar^2} \right ) ^{1/2} \sum_n \, (\mu - \epsilon_{ni})^{3/2} \, \Theta (\mu - \epsilon_{ni})\; .
\end{eqnarray}
We define
\begin{equation}
   \epsilon_{ni} = \epsilon_i + \bigl ( n + 1/2 \bigr ) \, \hbar \omega_{ic} + \gamma_i s \>,
\end{equation}
$\ell = (\hbar c / eB)^{1/2}$ is the magnetic length, and $\Theta(x)$ is the Heaviside function.

It is convenient to rescale lengths in units of the magnetic length $\ell$ and energies in units of the cyclotron energy of one of the pockets $\hbar \omega_{1c}$. Thence, in dimensionless units, we have
\begin{eqnarray}
\label{eq:n}
   n_i &=& \frac{\sqrt{2}}{ \pi^2} \left( \frac{m_{iz}}{m_{1c}} \right ) ^{1/2} \sum_n \, (\mu - \epsilon_{ni})^{1/2} \, \Theta (\mu - \epsilon_{ni}) \; ,\\
\label{eq:E}
   E_i &=& \frac{\sqrt{2}}{3 \pi^2} \left( \frac{m_{iz}}{m_{1c}} \right ) ^{1/2} \sum_n \, (\mu - \epsilon_{ni})^{3/2} \, \Theta (\mu - \epsilon_{ni}) \; .
\end{eqnarray}
The total energy of the system includes the strain energy and the electronic terms
\begin{equation}
\label{eq:totalen}
   U = \frac{1}{2} \, K s^2 + \sum_{i={1,2}} E_i \;\;.
\end{equation}


We are interested in following the total energy as a function of strain and magnetic field (which acts as a tuning parameter). To determine this, Eq.~\eqref{eq:n} fixes the chemical potential, fed in to determine the total energy via Eq.~\eqref{eq:E}. In general this is not analytically tractable except in special cases (e.g. particle hole symmetry, when $\mu =0$).

However, it is clear that the contribution of strain to the electronic energy is generally small, and therefore it is appropriate to separate out the singular contributions that occur near where Landau levels cross through the chemical potential from a smooth background, viz.
\begin{eqnarray}
\label{eq:singen}
   U & = & \frac{1}{2} \, K s^2 + U_{smooth}(s) \\ \nonumber
   && + \frac{\sqrt{2}\gamma^{3/2}}{3 \pi^2} \left ( \frac{m_z}{m_c} \right ) ^{1/2} (s_0-s)^{3/2} \, \Theta (s_0-s) \;,
\end{eqnarray}
where now
\begin{equation}
   s_0 = \frac{1}{\gamma} \frac{\Delta B}{B} \>,
\end{equation}
is the tuning parameter (magnetic field), which vanishes when the $n^\mathrm{th}$ Landau level of interest empties: $\Delta B = B_n - B$. (We are thus confined to Case I, for the moment.) The smoothly varying part of the energy can be incorporated into regular magnetostriction that will lead to a small spontaneous strain, that weakly renormalizes the effective stiffness, $K_\mathrm{eff}$, and shifts the parameter $s_0$. We re-parameterize to take this smooth variation into account. (And parenthetically we note that other strain-dependent parameters, such as  the effective mass, may be taken into account in the same way.) The last term is non-analytic, and needs to be treated separately from the background variation.

\begin{figure}[b]
   \centering
   \includegraphics[width=0.85\columnwidth]{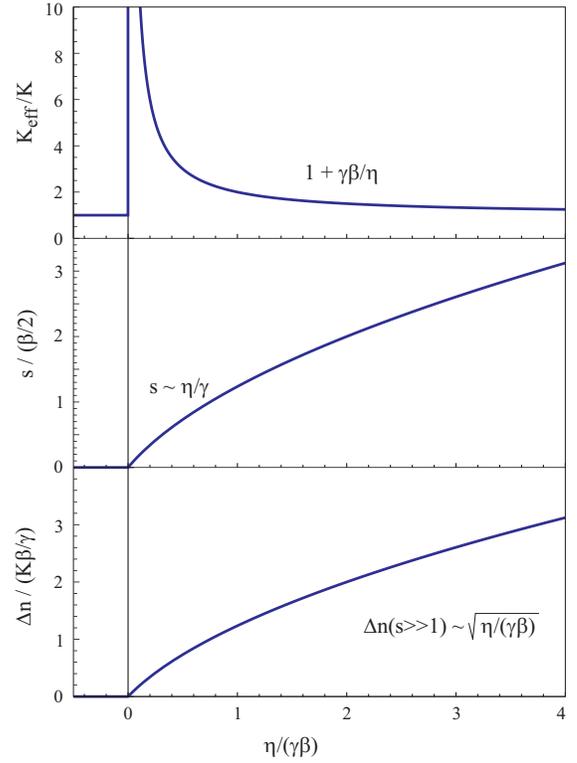}
\caption{\label{fig:sketch} (Color online) The effective elastic modulus from Eq.~\eqref{eq:stiffness}, induced strain from Eq.~\eqref{eq:strain}, and Landau level occupancy from \textcolor{black}{Eq.~\eqref{eq:occup}}, as a function of $\eta = \Delta B / B$. The latter is also proportional to the torque on the sample, when there is angular variation of $B_n (\theta)$.}
\end{figure}

By straightforward minimization, one finds
\begin{eqnarray}
   s = \left \{
   \begin{array}{l}
      0 \>,  \qquad {\rm for } \; \eta < 0 \>, \\
      \displaystyle
      \frac{\beta}{2} \left [ \left ( 1 + \frac{4 \eta}{\gamma \beta} \right ) ^{1/2} - 1 \right ]
      \simeq \frac{\eta}{\gamma} - \frac {\eta^2}{\beta \gamma } + \cdots
      \>,\\
      \qquad \quad {\rm for } \; \eta \geq 0 \>,
   \end{array}
   \right .
\label{eq:strain}
\end{eqnarray}
where we introduced the parameters $\eta = \Delta B / B$ and
\begin{equation}
   \beta = \frac{\gamma^3}{2 \pi^4 K^2} \, \frac{m_z}{m_c}
   \>.
\end{equation}
Corresponding to the kink in the spontaneous strain, there is a singular contribution to the effective stiffness, i.e.
\begin{equation}
\label{eq:stiffness}
   K_\mathrm{eff} = \frac{\partial ^2 U}{\partial s^2}
   = K \left [ 1 + \frac{\gamma \beta}{\eta} \, \Theta (s_0) \right ]
\end{equation}
After minimization, the internal energy becomes
\begin{eqnarray}
\label{eq:energy}
  U & = & \frac{K \beta^2}{12} \left [ \left ( 1 + \frac {4 \eta}{\gamma \beta} \right ) ^{3/2} - 1 - \frac{6 \eta}{\gamma \beta}  \right ]
  \\ \nonumber &
  \simeq & \frac{K\eta^2}{2\gamma^2} \left [ 1 - \frac{2\eta}{3\gamma \beta} + \cdots \right ] \>,
\end{eqnarray}
and the Landau level occupancy is given by
\begin{eqnarray}
\label{eq:occup}
 \Delta n
  =
  2 \, \frac{\partial U}{\partial \eta}
  & = & \frac{K \beta}{\gamma} \left [ - 1 + \Bigl ( 1 + 4 \, \frac {\eta}{\gamma \beta} \Bigr )^{1/2} \right ]
  \\ \nonumber &
  \simeq &
  2 \, \frac{K \beta}{\gamma} \left [ \frac{\eta}{\gamma \beta} - \Bigl ( \frac{\eta}{\gamma \beta} \Bigr )^2 + \cdots \right ] \>.
\end{eqnarray}
For large strains, the above approaches the Landau level occupancy result obtained when the induced strain is ignored, i.e.
\begin{eqnarray}
\label{eq:occup_s0}
  \Delta n_{s\equiv 0}
  & = &
  2 \, \frac{K \beta}{\gamma} \sqrt{ \frac{\eta}{\gamma \beta} }
  \ = \
  \frac{\sqrt{2}}{\pi^2} \ \sqrt{ \frac{m_z}{m_c}} \ \sqrt{\eta}
  \>,
\end{eqnarray}
where the $K$ and $\gamma$ dependences cancel out.
The size of the effects induced by strain is controlled by the small dimensionless parameter $\beta$, which is (re-dimensionalizing the other parameters)
\begin{equation}
   \beta = \frac{\gamma^3}{\pi^4 \hbar \omega_i K^2 \ell^6} \, \frac{m_z}{m_c}
   \>.
\end{equation}
The range in reduced field where the elastic corrections are observed is of order
$\Delta \eta = \gamma \beta /(\hbar \omega_c)$.

The above results are depicted in Fig.~\ref{fig:sketch} and represent the generic case of non-degenerate pockets (i.e. Case I). Case II, where there are degenerate pockets, is of interest in case the mechanism predicts an instability, breaking the symmetry between the pockets. By inspection, it is clear that it does not, because the energy increases as a power of the tuning parameter with an exponent greater than unity (3/2). Were the exponent less than unity, there would be a discontinuous jump in density.

but in dimensions less than 2 an instability is expected.


When the magnetic field is not aligned with a principal axis of the pockets, there is a torque, $\tau = \partial U / \partial \theta$. At the Landau level edge this acquires a discontinuous contribution, i.e.
\begin{equation}
   \tau = \frac{\partial U}{\partial \eta} \, \frac{\partial \log B_n}{\partial \theta}
   \>.
\end{equation}
Note that there is a contribution to the torque in the absence of coupling to strain, owing to the diamagnetic contribution of the carriers in the pockets. In the case considered here, the contribution at small excess fields arises because of the physical transfer of carriers from one pocket to another: the strain coupling {\em reduces} the torque and changes the behavior from square root to linear.


We now turn to a discussion of elemental \textit{Bi}. \textit{Bi} is a classic semimetal, having a hole pocket around the L-point on the $(111)$ surface of the Brillouin zone, and three electron pockets around the T-points on the $(\bar 1 11), (1 \bar 1 1), (1 1 \bar 1 )$ directions. One may interpret the band structure by realizing that the rhombohedral structure of \textit{Bi} is derived from a distorted simple cubic structure: there is a dimerization along the (111) direction, which forms a puckered sheet (though the dimerisation is small, with a difference of bond to back-bond lengths of 12\% ) and opens small gaps at the fermi energy; this is accompanied by a rhombohedral distortion from the ideal angle of $60^\circ$ to $57.2^\circ$ which breaks the symmetry between the L and T points on the Brillouin zone surface, leading to band overlap and a semimetal. Consequently, the band edges of the conduction and valence points are expected to be very sensitive to rhombohedral strain oriented along $(111)$; orthorhombic strain will break the symmetry of the electron pockets.

Since the early work of Kapitza and Shoenberg, \textit{Bi} has long been known to have very large magnetostriction~\cite{kapitza,shoenberg}, and the effects of quantizing magnetic fields have been studied by several authors~\cite{chandrasekhar,fawcett,michenaud}, with Michenaud \emph{et al}.~\cite{michenaud} presenting data up to 19~T and an analysis similar to the one above.

Recently, a number of studies have explored high-field ($>\!20$T) characteristics of Bi, though not explicitly the magnetoelastic effects. Nernst and Hall measurements~\cite{behnia07} revealed unexpected structure attributed to the hole pocket at fields large enough that the chemical potential is expected to be inside the lowest Landau level. Torque measurements as a function of angle (which are sensitive to the electron pockets) revealed sharp (and hysteretic) structure with strong angular dependence~\cite{ong}; subsequently the position (field, angle) of these anomalies has been identified~\textcolor{black}{\cite{behnia09,balents}} with electron pocket Landau level structure. The physics of this anomaly remains unidentified, but is consistent with breaking of the electron valley degeneracy. An alternative explanation~\textcolor{black}{\cite{fisher}} was proposed to explain the coincident observation of Hall plateaus at certain magic angles.

We focus on estimating the order of magnitude of the coupling the soft ``$c_{44}$'' rhombohedral shear, which is approximately the longitudinal strain along the trigonal axis. We calculate the electron-phonon coupling from first principles by varying the rhombohedral angle and calculating~\cite{details} the energies of the band edges at the L and T points, yielding $\gamma \approx 10 \pm 1 \; {\rm eV}$.
At 20~T, we estimate $\beta \approx 10^{-5}$,  consistent with the numerical estimates of Michenaud {\em et al.}~\cite{michenaud} using parameters extracted by Walther~\cite{walther}, and with their measurements at fields up to 19~T. Though the induced strains are small, they are significant.
The range in reduced field where the corrections are seen is $\beta \gamma / \hbar \omega_c \approx 0.1-0.5$, so that the conventional formulae for $n(\mu)$, $E(\mu)$ are invalid; the modulation in the pocket density is of order $K \beta / \gamma$, about $10^{15}$~cm$^{-3}$ at 20~T using the numbers above -- around 1\% of the zero-field carrier density.


We thus remark that the magnetoelastic effects in \textit{Bi} are expected to be unusually large, especially at high fields. As well as the predictions of ``single-particle'' physics, the sensitivity of strain to Landau-level occupancy would make such experiments useful to study interaction-driven transitions in this system, should they exist. Preliminary high-field measurements  show indeed substantial magneto-elastic quantum oscillations at fields exceeding 20 T~(see Ref.~\onlinecite{riggs}). The importance of such measurements is that they are sensitive to the total electron and hole numbers in the pockets, and are intrinsically easier to interpret than transport data. We predict a singularity in the sound velocity at the Landau-level edge in a clean system.

With the exception of the high field Nernst measurements~\cite{behnia07} on the fractionally filled hole pocket, there is as yet no strong evidence for correlation effects in \textit{Bi} at other than close to integer Landau level fillings. Nonetheless, there are several puzzles in the data that have been taken as evidence for the modulation of carrier densities in the valleys. It is important to distinguish between effects that indicate a spontaneous instability of the system from those that can be derived mechanically.

Consequently, the electron Landau-level lines that rise steeply at high fields, that were identified by the field $H_2 (\theta)$ (Fig.~3 of Ref.~\onlinecite{ong}) in torque measurements, are not special for their existence (which follows the most recent~\cite{balents,fisher}, though not earlier calculations), but for their shape. The observed hysteresis near zero angle is not compatible with single-particle effects as discussed here, though the approximate shape of the anomaly is consistent with Fig.~\ref{fig:sketch} for angles in the range of 2-4 degrees where no hysteresis is reported.

The observed (see Fig.~8 of Ref.~\onlinecite{behnia09}) reversal of the sign of the torque at magic angles for rotations in the trigonal-bisectrix plane is consistent with the crossing of (different) Landau levels of the doubly-degenerate and singly-degenerate electron pockets, with the sign-change arising from the opposite angular-dependence of the Landau level shifts. Associated with this effect, it was noted that the Hall resistance $\rho_{xy}$ -- in this geometry believed to be dominated by the holes -- is field-independent over a range of several Tesla. In contrast, for rotations in the trigonal-binary plane (when only a single electron pocket is involved), the Hall resistance shows a pronounced angular minimum but continues to grow with field. We note that the magnetoelastic modulation of the carrier density due to rhombohedral strain will modulate electron and hole densities equally; this is a mechanism for the electron Landau-level crossings to moderate transport by the holes, aside from any direct contribution of the electron carriers to transport coefficients. Since $\Delta n \propto \tau$, one may speculate that the near cancelation of the torque anomalies along the trigonal-bisectrix axis might be reflected in $\rho_{xy}$, in contrast to the single Landau level physics in the binary direction.

Finally, we must make a distinction between the elastic strain induced effects discussed here (at long wavelengths, $q=0$) and the possibility of finite$-q$ phonon softening and instabilities due to field-induced nesting, much studied in the context of field-induced charge and spin density waves~\cite{brazovskii82}. At zero magnetic field, the Bi electron and hole pockets do not nest, and so a density-wave instability is suppressed. At large fields, the reduced dimensionality of the dispersion will give rise to one-dimensional-like nesting features and hence an instability. Note, however, that (at least in weak coupling) the transition temperature will generically scale with a coupling constant $\lambda$ as $T_c \propto e^{-1/\lambda}$, where $\lambda \propto m_{iz}$ via its dependence on the density of states. The very light mass of carriers in Bi drives such an instability to very low temperature.


In conclusion, we have made a quantitative estimate of magnetoelastic coupling effects in a degenerate semimetal in the quantum limit, predicting the detailed shape of anomalies that should be measurable in \textit{Bi}. In contrast to two dimensions, we find that electron-phonon effects {\em per se} do not produce instabilities, but lead to discontinuities in derivatives of the elastic strain and a singularity in the modulus.

We believe that carrying out such measurements in \textit{Bi} at high fields -- and preliminary data has been obtained~\cite{riggs} -- would be very helpful in unraveling the complex physics that has emerged from other recent experiments~\cite{behnia07,ong,behnia09} on this system.

\begin{acknowledgements}
PBL thanks Los Alamos National Laboratory and the National High Magnetic Field Laboratory for hospitality during this work. We are grateful to S. Riggs for discussions and for the sight of preliminary data. The authors gratefully acknowledge useful conversations with D.L. Smith. This work was supported in part by the Engineering and Physical Sciences Research Council (EPSRC) and the Los Alamos National Laboratory under the auspices of the U.S. Department of Energy.
\end{acknowledgements}

\end{document}